\documentclass[10pt]{bmc_article}

\usepackage{cite} 
\usepackage{url}  
\usepackage{ifthen}  
\usepackage{multicol}   
\usepackage[utf8]{inputenc} 
\usepackage{graphicx}
\usepackage{graphics}
\usepackage[normalem]{ulem}
\usepackage{color}

\newboolean{publ}

\newenvironment{bmcformat}{\begin{raggedright}\baselineskip20pt\sloppy\setboolean{publ}{false}}{\end{raggedright}\baselineskip20pt\sloppy}

\begin{document}
\begin{bmcformat}

\title{abc: an R package for Approximate Bayesian Computation (ABC)}
 
\author{Katalin Csill\'ery\correspondingauthor$^{1,2}$%
  \email{Katalin Csill\'ery\correspondingauthor - kati.csillery@gmail.com}%
  \and
  Olivier Fran\c cois$^1$%
  \email{Olivier Fran\c cois - olivier.francois@imag.fr}%
  \and  
  and
  Michael GB Blum\correspondingauthor$^1$%
  \email{Michael GB Blum\correspondingauthor - michael.blum@imag.fr}%
}

\address{%
  \iid(1) Universit\' e Joseph Fourier, Grenoble 1, Centre National de la Recherche Scientifique, Laboratoire TIMC-IMAG UMR 5525, Equipe Biologie Computationnelle et Math\'ematique, Grenoble, F-38041, France
  \iid(2) CEMAGREF, UR Ecosyst\`emes montagnards, 2 rue de la papeterie, 38402 Saint Martin d'H\`eres, France
}

\maketitle

\begin{abstract}
  \paragraph*{Background:} 
  Many recent statistical applications involve inference under complex
  models, where it is computationally prohibitive to calculate
  likelihoods but possible to simulate data.  Approximate Bayesian
  Computation (ABC) is devoted to these complex models because it
  bypasses evaluations of the likelihood function using comparisons
  between observed and simulated summary statistics.
  \paragraph*{Results:}
  We introduce the \verb|R| \textbf{abc} package that implements
  several ABC algorithms for performing parameter estimation and model
  selection. In particular, the recently developed non-linear
  heteroscedastic regression methods for ABC are implemented. The
  \textbf{abc} package also includes a cross-validation tool for
  measuring the accuracy of ABC estimates, and to calculate the
  misclassification probabilities when performing model selection. The main
  functions are accompanied by appropriate summary and plotting
  tools. Considering an example of demographic inference with
  population genetics data, we show the potential of the \verb|R|
  package.
  \paragraph*{Conclusions:}
  \verb|R| is already widely used in bioinformatics and several fields
  of biology. The \verb|R| \textbf{abc} package will make the ABC
  algorithms available to the large number of \verb|R|
  users. \textbf{abc} is a freely available \verb|R| package under the
  GPL license, and it can be downloaded at
  \url{http://cran.r-project.org/web/packages/abc/index.html}.
\end{abstract}

\ifthenelse{\boolean{publ}}{\begin{multicols}{2}}{}


\section*{Background}

In recent years, Approximate Bayesian computation (ABC) has become a
popular method for parameter inference and model selection under
complex models, where the evaluation of the likelihood function is
computationally prohibitive.  ABC bypasses exact likelihood
calculations via the use of summary statistics and simulations, which,
in turn, allows the consideration of highly complex models. ABC was
first proposed in population genetics to do inference under coalescent
models \cite{beaumontetal02,pritchardetal99}, but it is now
increasingly applied in other fields, such as ecology or systems
biology (see \cite{beaumont10,bertorelleetal10,csilleryetal10} for
reviews of ABC methods and applications). Software implementations of
ABC dedicated to particular problems have already been developed in
these fields \cite{andersonetal05,cornuetetal08,tallmonetal08,lopesetal09,
  thornton09, cornuetetal10, liepeetal10, wegmannetal10, brayetal10,
  hickersonetal07, wenetal11, rejector} (See additional file : table giving the main features of the ABC software).\pb

The integration of ABC in a software package poses several
challenges. First, data simulation, which is in the core of any ABC
analysis, is specific to the model in question. Thus, many existing
ABC software are specific to a particular class of models
\cite{hickersonetal07, cornuetetal08, lopesetal09} or even to the
estimation of a particular parameter \cite{tallmonetal08}. Further,
model comparison is an integral part of any Bayesian analysis, thus it
is essential to provide software, where users are able to fit
different models to their data. Second, an ABC analysis often follows
trial-error approaches, where users try different models, ABC
algorithms, or summary statistics. Therefore, it is important that
users can run different analyses using batch files, which contain each
analysis as a sequence of commands.  Third, ABC is subject to
intensive research and many new algorithms have been published in the
past few years \cite{bortotetal07, sissonetal07, beaumontetal09,
  blum10}. Thus, an ABC software should be flexible enough to
accommodate new developments in the field.\pb

Here, we introduce a generalist \verb|R| package, \textbf{abc}, which
aims to address the above challenges \cite{R10}. The price to pay for
the generality and flexibility is that the simulation of data and the
calculation of summary statistics are left to the users. However,
simulation software might be called from an \verb|R| session, which
opens up the possibility for a highly interactive ABC analysis. For
coalescent models, for instance, users can apply one of the many
existing software for simulating genetic data such as \verb|ms|
\cite{hudson02} or \verb|simcoal2| \cite{simcoal2}. \verb|R| provides
many advantages in the context of ABC: (i) \verb|R| already possesses
the necessary tools to handle, analyze and visualize large data sets,
(ii) sequences of \verb|R| commands can be saved in a script file,
(iii) \verb|R| is a free and collaborative project, thus new
algorithms can be easily integrated to the package (e.g. via
contributions from their authors). \pb

\section*{Implementation}

The main steps of an ABC analysis follow the general scheme of any
Bayesian analysis: formulating a model, fitting the model to data
(parameter estimation), and improving the model by checking its fit
(posterior predictive checks) and comparing it to other models
\cite{gelmanetal03, csilleryetal10}. \textbf{abc} provides functions
for the inference and model comparison steps, and we indicate how
generic tools of the \verb|R-base| package can be used for model
checking.\pb

In order to use the package the following data should be prepared for
\verb|R|: a vector of the observed summary statistics, a matrix of the
simulated summary statistics, where each row corresponds to a
simulation and each column corresponds to a summary statistic, and
finally, matrix of the simulated parameter values, where each row
corresponds to a simulation and each column corresponds to a
parameter.

\subsection*{Parameter inference}

For the sake of clarity, we recall the general scheme of parameter
estimation with ABC. Suppose that we want to compute the posterior
probability distribution of a univariate or multivariate parameter,
$\theta$.  A parameter value $\theta_i$, is sampled from its prior
distribution to simulate a dataset $y_i$, for $i=1,\dots, n$ where $n$
is the number of simulations. A set of summary statistics $S(y_i)$ is
computed from the simulated data and compared to the summary
statistics obtained from the actual data $S(y_0)$ using a distance
measure $d$. If the distance $d(S(y_i),S(y_0))$ is less than a given
threshold, the parameter value $\theta_i$ is accepted.  The accepted
$\theta_i$'s form a sample from an approximation of the posterior
distribution, which can be improved by the use of regression
techniques (see below). \pb

The function \verb|abc| implements three ABC algorithms for
constructing the posterior distribution from the accepted
$\theta_i$'s: a rejection method, and regression-based correction
methods that use either local linear regression \cite{beaumontetal02}
or neural networks \cite{blumfrancois10}. When the rejection method
(\verb|"rejection"|) is selected, the accepted $\theta_i$'s are
considered as a sample from the posterior distribution
\cite{pritchardetal99}. The two regression methods (\verb|"loclinear"|
and \verb|"neuralnet"|) implement an additional step to correct for
the imperfect match between the accepted, $S(y_i)$, and observed
summary statistics, $S(y_0)$, using the following regression equation
in the vicinity of $S(y_0)$
$$
\theta_i=m(S(y_i))+ \epsilon_i,
$$ where $m$ is the regression function, and the $\epsilon_i$'s are
centered random variables with a common variance. Simulations that
closely match $S(y_0)$ are given more weight by assigning to each
simulation $(\theta_i,S(y_i))$ the weight $K[d(S(y_i),S(y_0))]$, where
$K$ is a statistical kernel. The local linear model
(\verb|"loclinear"|) assumes a linear function for $m$, while neural
networks account for the non-linearity of $m$ and allow users to
reduce the dimension of the set of summary statistics. Once the
regression is performed, a weighted sample from the posterior
distribution is obtained by correcting the $\theta_i$'s as follows,
$$
\theta_i^*=\hat{m}(S(y_0))+\hat{\epsilon_i},
$$ where $\hat{m}(\cdot)$ is the estimated conditional mean and the
$\hat{\epsilon_i}$'s are the empirical residuals of the regression
\cite{beaumontetal02}. Additionally, a correction for
heteroscedasticity is applied, by default, in \verb|abc|,
$$
\theta_i^*=\hat{m}(S(y_0))+\frac{\hat{\sigma}(S(y_0))}{\hat{\sigma}(S(y_i))}
\hat{\epsilon_i}
$$ where $\hat{\sigma}(\cdot)$ is the estimated conditional standard
deviation \cite{blumfrancois10}.\pb

The function \verb|abc| returns an object of class \verb| "abc"| that
can be printed, summarized and plotted using the S3 methods of the
\verb|R| generic functions, \verb|print|, \verb|summary|, \verb|hist|
and \verb|plot|. The function \verb|print| returns a summary of
the object. The function \verb|summary| calculates summaries of the
posterior distributions, such as the mode, mean, median, and credible
intervals, taking into account the posterior weights, when
appropriate. The \verb|hist| function displays the histogram of the
weighted posterior sample. The \verb|plot| function generates various
plots that allow the evaluation of the quality of estimation when one
of the regression methods is used. The following plots are generated:
a density plot of the prior distribution, a density plot of the
posterior distribution estimated with and without regression-based
correction, a scatter plot of the Euclidean distances as a function of
the parameter values, and a normal Q-Q plot of the residuals from the
regression. When the heteroscedastic regression model is used, a
normal Q-Q plot of the standardized residuals is displayed.\pb

Finally, we note that alternative algorithms exist that sample from an
updated distribution that is closer in shape to the posterior than to
the prior \cite{beaumontatal09, marjorametal03,
  wegmannetal10}. However, we do not implement these methods in
\verb|R| \textbf{abc} because they require the repeated use of the
simulation software.\pb

\subsection*{Cross-validation}

The function \verb|cv4abc| performs a leave-one-out
cross-validation to evaluate the accuracy of parameter estimates and
the robustness of the estimates to the tolerance level. To perform
cross-validation, the $i^{th}$ simulation is randomly selected as a
validation simulation, its summary statistic(s) $S(y_i)$ are used as
``dummy'' observed summary statistics, and its parameters are
estimated via \verb|abc| using all simulations except the $i^{th}$
simulation. Ideally, the process is repeated $n$ times, where $n$ is
the number of simulations (so-called $n$-fold
cross-validation). However, performing an $n$-fold cross-validation
might take up too much time, so the cross-validation is often
performed for a subset of typically 100 randomly selected simulations. The
\verb|summary| S3 method of \verb|cv4abc| computes the prediction
error as
$$ E_{\rm pred}=\frac{\sum_i(\tilde{\theta}_i-\theta_i)^2}{Var(\theta_i)},
$$ where $\theta_i$ is the true parameter value of the $i^{th}$
simulated data set and $\tilde{\theta}_i$ is the estimated parameter
value. The \verb|plot| function displays the estimated parameter values as a
function of the true values.\pb

\subsection*{Model selection}

The function \verb|postpr| implements model selection to estimate the
posterior probability of a model $M$ as $Pr(M | S(y_0))$. Three
different methods are implemented. With the rejection method
(\verb|"rejection"|), the posterior probability of a given model is
approximated by the proportion of accepted simulations given this
model. The two other methods are based on multinomial logistic
regression (\verb|"mnlogistic"|) or neural networks
(\verb|"neuralnet"|). In these two approaches, the model indicator is
treated as the response variable of a polychotomous regression, where
the summary statistics are the independent variables
\cite{beaumont08}. Using neural networks can be efficient when highly
dimensional statistics are used. Any of these methods are valid when the
different models to be compared are, a priori, equally likely, and the
same number of simulations are performed under each model. The
\verb|postpr|'s S3 method for \verb|summary| displays the posterior
model probabilities, and calculates the ratios of model probabilities,
the Bayes factor, for all possible pairs of models
\cite{francoisetal08}.\pb

A further function, \verb|expected.deviance|, is implemented to guide
the model selection procedure. The function computes an approximate
expected deviance from the posterior predictive distribution. Thus, in
order to use the function, users have to re-use the simulation tool
and to simulate data from the posterior parameter values. The
method is particularly advantageous when it is used with one of the
regression methods. Further details on the method can be found in
\cite{francoislaval11} and fully worked out examples are provided in
the package manual.

\subsection*{Model misclassification}

A cross-validation tool is available for model selection as well. The
objective is to evaluate if model selection with ABC is able to
distinguish between the proposed models by making use of the existing
simulations. The summary statistics from one of the simulations are
considered as pseudo-observed summary statistics and classified using
all the remaining simulations. Then, one expects that a large
posterior probability should be assigned to the model that generated
the presudo-observed summary statistics. Two versions of the
cross-validation are implemented. The first version is a ``hard''
model classification. We consider a given simulation as the
pseudo-observed data assign it to the model for which \verb|postpr|
gives the highest posterior model probability. This procedure is
repeated for a given number of simulations for each model. The results
are summarized in a so-called {\it confusion matrix}
\cite{hastieetal03}. Each row of the confusion matrix represents the
number of simulations under a true model, while each column represents
the number of simulations under a model assigned by \verb|postpr|. If
all simulations had been correctly classified, only the diagonal
elements of the matrix are non-zero. The second version is called
``soft'' classification. Here we do not assign a simulation to the
model with the highest posterior probability, but average the
posterior probabilities over many simulations for a given model. This
procedure is again summarized as a matrix, which is similar to the
confusion matrix. However, the elements of the matrix do not give
model counts, but the average posterior probabilities across
simulations for a given model. The matrices can be visualized using a
barplot using the S3 method for \verb|plot|.


\section*{Results and Discussion}

In following sections we show the utility of the \textbf{abc} package
using an example of parameter inference in human demographic history.

\subsection*{Background and data}
Several studies have found evidence that African human populations
have been expanding, while human populations outside of Africa went
through a population bottleneck and then expanded. We re-examined this
problem by re-analyzing published data of 50 unlinked autosomal
non-coding regions from a Hausa (Cameroon) population and a European
(Italy) population \cite{voightetal05}. Data were summarized using
three summary statistics: the mean and the variance of Tajima's D and
the mean heterozygosity (H) (Table 2 in \cite{voightetal05}). Both
Tajima's D and H have been classically used to detect historical
changes in population size. Here, we analyzed these data using
\textbf{abc} to provide a walk-through analysis of the functions of
the \verb|R| package. We reached conclusions in accordance with
\cite{voightetal05}, who analyzed these data with a different method
(see details in \cite{voightetal05}). We found support for the
hypothesis that the Hausa population is expanding whereas the Italian
population has experienced a bottleneck. Additionally, we estimated
the ancestral Italian population size to be around $10,000$
individuals, which is similar to the results of
\cite{voightetal05}.\pb

\subsection*{Model selection}
We considered three different models of demographic history: constant
population size, exponential growth after a period of constant
population size, and population bottleneck, where, after the
bottleneck, the population recovered to its original size. All three
models can be defined by a number of demographic parameters, such as
population sizes, rates and timings of changes in population size,
which we do not detail here. We simulated 50,000 data sets under each
of the three demographic models using the software \verb|ms|
\cite{hudson02}, and calculated the three summary statistics in each
model.

Before applying the model selection function (\verb|postpr|) on the
real data, we performed a cross-validation for model selection
(\verb|cv4postpr|) to evaluate if ABC can, at all, distinguish between
these three models. We performed 500 cross-validation simulations for
each model. The following confusion matrix (using the S3 method for
\verb|summary|) and the barplot (using the S3 method for \verb|plot|,
Figure 1) illustrate that ABC is able to distinguish between these
three models, and, notably, it is the bottleneck model that can be
classified correctly the most frequently: 396 times out of 500.
\begin{verbatim}
Confusion matrix based on 500 samples for each model.
$tol0.01
       bott const exp
bott    396    97   7
const    92   321  87
exp      34   144 322

Mean model posterior probabilities using the mnlogistic method.
$tol0.01
        bott  const    exp
bott  0.7202 0.2154 0.0644
const 0.2204 0.5085 0.2711
exp   0.0791 0.2960 0.6249
\end{verbatim} 

Next, we calculated the posterior probabilities of each demographic
scenario using the rejection (\verb|"rejection"|) and the multinomial
logistic regression method (\verb|"mnlogistic"|) of the function
\verb|postpr|. Considering three different tolerance rates, $0.1\%$,
$0.5\%$, and $1\%$, we found that the Hausa data best supported the
model of exponential growth whereas the Italian data best supported
the bottleneck model (see Table 1). Application of the \verb|summary|
method of \verb|postpr| produced the following output for the Italian
data:
\begin{verbatim}
Call: 
postpr(target = tajima.obs$italy, index = models, sumstat =
 tajima.sim, tol = 0.1, method = "mnlogistic")
Data:
 postpr.out$values (15000 posterior samples)
Models a priori:
 bott, const, exp
Models a posteriori:
 bott, const, exp

Proportion of accepted simulations (rejection):
 bott const   exp 
0.873 0.113 0.014 

Bayes factors:
        bott  const    exp
bott   1.000  7.754 61.201
const  0.129  1.000  7.893
exp    0.016  0.127  1.000

Posterior model probabilities (mnlogistic):
 bott const   exp 
0.991 0.009 0.001 

Bayes factors:
          bott    const      exp
bott     1.000  113.154 1745.512
const    0.009    1.000   15.426
exp      0.001    0.065    1.000
\end{verbatim}

\subsection*{Posterior predictive checks}

To further confirm which model provided the best fit to the data, we
considered posterior predictive checks \cite{gelmanetal03}. Note that
there is no specific function in \textbf{abc} for posterior predictive
checks, nevertheless the task can be easily carried out using the
simulation software (here \verb|ms|) and \verb|R|. Here, we illustrate
how the posterior predictive checks were run for the Italian
sample.\pb

First, we estimated the posterior distributions of the parameters of
the three models using \verb|abc|. Then, we sampled a set of 1,000
multivariate parameters from their posterior distribution. Last, we
obtained a sample from the distribution of the three summary
statistics a posteriori by simulating data sets with the 1,000 sampled
multivariate parameters using \verb|ms| (see Figure 2).  We found that
the model of exponential growth was unable to account for the observed
level of heterozygosity when accounting for the remaining two summary
statistics in the Italian sample. The bottleneck and the constant-size
population were, however, able to reproduce the observed values of the
summary statistics. Such posterior predictive checks use the summary
statistics twice; once for sampling from the posterior and once for
comparing the marginal posterior predictive distributions to the
observed values of the summary statistics. An alternative approach of
{\it model criticism} use the summary statistics only once but is
based on ABC-MCMC, which is not implemented in the package
\cite{ratmannetal09}.\pb

\subsection*{Parameter inference}
Next we inferred the parameters of the bottleneck model for the
Italian data. The bottleneck model can be described with four
parameters: the ancestral population size $N_a$, the ratio of the
population sizes before and during the bottleneck, the duration of the
bottleneck, and the time since the beginning of the bottleneck. We
show only the estimation of the ancestral population size $N_a$. We
first assessed if ABC was able to estimate the parameter $N_a$ at
all. We used the function \verb|cv4abc| to determine the accuracy of
ABC and the sensitivity of estimates to the tolerance rate. Figure 3
shows the estimated values of $N_a$ as a function of the true values
for three distinct tolerance rates. The posterior distribution of
$N_a$ was summarized with its median. The points of the
cross-validation plot are scattered around the identity line
indicating that $N_a$ can be well estimated using the three summary
statistics. Further, estimates were not only accurate for $N_a$, but
also insensitive to the tolerance rate (see Figure 3). Accordingly,
the prediction error was found to be around $0.20$ independently of
the tolerance rate.\pb

When using \verb|abc| with the method \verb|"neuralnet"|, we obtained
the following summary of the posterior distribution using
the function \verb|summary|:
\begin{verbatim}
Data:
 abc.out$adj.values (500 posterior samples)
Weights:
 abc.out$weights
                        Ancestral_Size
Min.:                   7107.099
Weighted 2.5 % Perc.:   7604.957
Weighted Median:        9951.219
Weighted Mean:         10148.146
Weighted Mode:         10004.551
Weighted 97.5 % Perc.: 14284.484
Max.:                  16505.348
\end{verbatim}

The function \verb|plot| can be used as a diagnostic tool for the
visualization of the output of \verb|abc| for the parameter
$N_a$. Figure 4 shows that the posterior distribution is very
different from the prior distribution, confirming that the three
summary statistics convey information about the ancestral population
size (Figure 4 lower left panel). The upper right panel of Figure 4
shows the distance between the simulated and observed summary
statistics as a function of the prior values of $N_a$. Again, this
plot confirms that the summary statistics convey information about
$N_a$, because the distances corresponding to the accepted values are
clustered and not spread around the prior range of $N_a$. The lower
right panel displays a standard Q-Q plot of the residuals of the
regression, $\hat{\epsilon_i}$. This plot serves as a regression
diagnostic of \verb|locfit| or \verb|nnet|, when the method is
\verb|"loclinear"| or \verb|"neuralnet"|.\pb


\section*{Conclusions}
We provide an \verb|R| package \textbf{abc} to perform model selection
and parameter estimation via Approximate Bayesian
Computation. Integrating \textbf{abc} within the \verb|R| statistical
environment offers high quality graphics and data visualization
tools. The \verb|R| package implements recently developed non-linear
methods for ABC, and is going to evolve as new algorithms and methods
accumulate.\pb

\section*{Availability and requirements} 

\textbf{Project name:} \verb|R| \textbf{abc}\\
\textbf{Project home page}: \url{http://cran.r-project.org/web/packages/abc/index.html}\\
\textbf{Version:} 1.3\\
\textbf{Operating system(s):} Windows, Linux, MacOS\\
\textbf{Programming language:} \verb|R|\\
\textbf{Other requirements:} \verb|R| version $\geq$2.10 and the \verb|R| packages: \textbf{nnet, quantreg, locfit}\\
\textbf{License:} GNU GPL$\geq$3\\
\textbf{Any restrictions to use by non-academics:} None\\

\section*{Authors' contributions}
    MB and KC implemented the algorithms and methods, and wrote the
    paper. KC created the \verb|R| package. OF critically reviewed the paper
    and tested the package.

\section*{Acknowledgements and Funding}
  \ifthenelse{\boolean{publ}}{\small}{} We thank Mark Beaumont for
  kindly providing an \verb|R| script that we used in the
  implementation of the functions \verb|abc| and \verb|postpr|. While
  working on this package, KC was funded by a post-doctoral fellowship
  from the Universit\'e Joseph Fourier (ABC MSTIC), and then was
  hosted and financed by the Ecology and Evolution Laboratory (ENS,
  Paris, ANR-06-BDIV-003).
  
 \section*{Additional File}
  \ifthenelse{\boolean{publ}}{\small}{} 
  File name: ABCtable.pdf\\
  File format: .pdf\\
  Title of file: Main features of the ABC software\\
  Description of file: Table describing the main features of the ABC software.

{\ifthenelse{\boolean{publ}}{\footnotesize}{\small}
 \bibliographystyle{bmc_article}  
  \bibliography{csilleryetal} }     


\begin{thebibliography}{10}
\providecommand{\url}[1]{[#1]}
\providecommand{\urlprefix}{}

\bibitem{beaumontetal02}
Beaumont MA, Zhang W, Balding DJ: \textbf{Approximate {B}ayesian computation in
  population genetics}. \emph{Genetics} 2002, \textbf{162}:2025--2035.

\bibitem{pritchardetal99}
Pritchard JK, Seielstad MT, Perez-Lezaun A, Feldman MW: \textbf{Population
  growth of human {Y} chromosomes: a study of Y chromosome microsatellites}.
  \emph{Molecular Biology and Evolution} 1999, \textbf{16}:1791--1798.

\bibitem{beaumont10}
Beaumont MA: \textbf{Approximate Bayesian Computation in Evolution and
  Ecology}. \emph{Annual Review of Ecology, Evolution, and Systematics} 2010,
  \textbf{41}:379--406.

\bibitem{bertorelleetal10}
Bertorelle G, Benazzo A, Mona S: \textbf{ABC as a flexible framework to
  estimate demography over space and time: some cons, many pros}.
  \emph{Molecular Ecology} 2010, \textbf{19}:2609--2625.

\bibitem{csilleryetal10}
Csill\'ery K, Blum MGB, Gaggiotti OE, Fran{\c c}ois O: \textbf{{Approximate
  Bayesian Computation in practice}}. \emph{Trends Ecol Evol} 2010,
  \textbf{25}:410--418.

\bibitem{andersonetal05}
Anderson CNK, Ramakrishnan U, Chan YL, Hadly EA: \textbf{Serial SimCoal: A
  population genetics model for data from multiple populations and points in
  time}. \emph{Bioinformatics} 2005, \textbf{21}(8):1733--1734,
  \urlprefix\url{[http://bioinformatics.oxfordjournals.org/content/21/8/1733.abstract]}.

\bibitem{cornuetetal08}
Cornuet JM, Santos F, Beaumont MA, Robert CP, Marin JM, Balding DJ, Guillemaud
  T, Estoup A: \textbf{{Inferring population history with DIY ABC: a
  user-friendly approach to approximate Bayesian computation}}.
  \emph{Bioinformatics} 2008, \textbf{24}(23):2713--2719,
  \urlprefix\url{[http://bioinformatics.oxfordjournals.org/content/24/23/2713.abstract]}.

\bibitem{tallmonetal08}
Tallmon DA, Koyuk A, Luikart G, Beaumont MA: \textbf{COMPUTER PROGRAMS:
  onesamp: a program to estimate effective population size using approximate
  Bayesian computation}. \emph{Molecular Ecology Resources} 2008,
  \textbf{8}(2):299--301,
  \urlprefix\url{[http://dx.doi.org/10.1111/j.1471-8286.2007.01997.x]}.

\bibitem{lopesetal09}
Lopes JS, Balding D, Beaumont MA: \textbf{{PopABC: a program to infer
  historical demographic parameters}}. \emph{Bioinformatics} 2009,
  \textbf{25}(20):2747--2749,
  \urlprefix\url{[http://bioinformatics.oxfordjournals.org/content/25/20/2747.abstract]}.

\bibitem{thornton09}
Thornton K: \textbf{Automating approximate Bayesian computation by local linear
  regression}. \emph{BMC Genetics} 2009, \textbf{10}:35,
  \urlprefix\url{[http://www.biomedcentral.com/1471-2156/10/35]}.

\bibitem{cornuetetal10}
Cornuet JM, Ravigne V, Estoup A: \textbf{Inference on population history and
  model checking using DNA sequence and microsatellite data with the software
  DIYABC (v1.0)}. \emph{BMC Bioinformatics} 2010, \textbf{11}:401,
  \urlprefix\url{[http://www.biomedcentral.com/1471-2105/11/401]}.

\bibitem{liepeetal10}
Liepe J, Barnes C, Cule E, Erguler K, Kirk P, Toni T, Stumpf MP:
  \textbf{{ABC-SysBio---approximate Bayesian computation in Python with GPU
  support}}. \emph{Bioinformatics} 2010, \textbf{26}(14):1797--1799,
  \urlprefix\url{[http://bioinformatics.oxfordjournals.org/content/26/14/1797.abstract]}.

\bibitem{wegmannetal10}
Wegmann D, Leuenberger C, Neuenschwander S, Excoffier L: \textbf{ABCtoolbox: a
  versatile toolkit for approximate Bayesian computations}. \emph{BMC
  Bioinformatics} 2010, \textbf{11}:116,
  \urlprefix\url{[http://www.biomedcentral.com/1471-2105/11/116]}.

\bibitem{brayetal10}
Bray TC, Sousa V, Parreira B, Bruford M, Chikhi L: \textbf{2BAD: an application
  to estimate the parental contributions during two independent admixture
  events}. \emph{Molecular Ecology Resources} 2010, \textbf{10}:538,541.

\bibitem{hickersonetal07}
Hickerson M, Stahl E, Takebayashi N: \textbf{msBayes: Pipeline for testing
  comparative phylogeographic histories using hierarchical approximate Bayesian
  computation}. \emph{BMC Bioinformatics} 2007, \textbf{8}:268,
  \urlprefix\url{[http://www.biomedcentral.com/1471-2105/8/268]}.

\bibitem{wenetal11}
Huang W, Takebayashi N, Qi Y, Hickerson M: \textbf{MTML-msBayes: Approximate
  Bayesian comparative phylogeographic inference from multiple taxa and
  multiple loci with rate heterogeneity}. \emph{BMC Bioinformatics} 2011,
  \textbf{12}:1, \urlprefix\url{[http://www.biomedcentral.com/1471-2105/12/1]}.

\bibitem{rejector}
Jobin M, Mountain J: \textbf{REJECTOR: software for population history
  inference from genetic data via a rejection algorithm}. \emph{Bioinformatics}
  2008, \textbf{24}:2936--2937.

\bibitem{bortotetal07}
Bortot P, Coles SG, Sisson SA: \textbf{Inference for stereological extremes}.
  \emph{Journal of the American Statistical Association} 2007,
  \textbf{102}:84--92.

\bibitem{sissonetal07}
Sisson SA, Fan Y, Tanaka M: \textbf{Sequential {M}onte {C}arlo without
  likelihoods}. \emph{Proceedings of the National Academy of Sciences of the
  United States of America} 2007, \textbf{104}:1760--1765. [Errata (2009), 106,
  16889].

\bibitem{beaumontetal09}
Beaumont MA, Marin JM, Cornuet JM, Robert CP: \textbf{Adaptivity for {A}{B}{C}
  algorithms: the {A}{B}{C}-{P}{M}{C} scheme}. \emph{Biometrika} 2009,
  \textbf{96}:983--990.

\bibitem{blum10}
Blum MGB: \textbf{Approximate Bayesian Computation: A Nonparametric
  Perspective}. \emph{Journal of the American Statistical Association} 2010,
  \textbf{105}(491):1178--1187,
  \urlprefix\url{[http://pubs.amstat.org/doi/abs/10.1198/jasa.2010.tm09448]}.

\bibitem{R10}
{R Development Core Team}: \emph{R: A Language and Environment for Statistical
  Computing}. R Foundation for Statistical Computing, Vienna, Austria 2010,
  \urlprefix\url{[http://www.R-project.org]}. [{ISBN} 3-900051-07-0].

\bibitem{hudson02}
Hudson RR: \textbf{{Generating samples under a Wright-Fisher neutral model of
  genetic variation}}. \emph{Bioinformatics} 2002, \textbf{18}(2):337--338.

\bibitem{simcoal2}
Laval G, Excoffier L: \textbf{{SIMCOAL} 2.0: a program to simulate genomic
  diversity over large recombining regions in a subdivided population with a
  complex history}. \emph{Bioinformatics} 2004, \textbf{20}(15):2485--2487.

\bibitem{gelmanetal03}
Gelman A, Carlin JB, Stern HS, Rubin DB: \emph{Bayesian Data Analysis, Second
  Edition (Texts in Statistical Science)}. Boca Raton: Chapman \& Hall/CRC, 2
  edition 2003.

\bibitem{blumfrancois10}
Blum MGB, Fran\c{c}ois O: \textbf{Non-linear regression models for
  {A}pproximate {B}ayesian {C}omputation}. \emph{Statistics and Computing}
  2010, \textbf{20}:63--73.

\bibitem{beaumontatal09}
Beaumont MA, Cornuet JM, Marin JM, Robert CP: \textbf{Adaptive approximate
  Bayesian computation}. \emph{Biometrika} 2009, \textbf{96}(4):983--990.

\bibitem{marjorametal03}
Marjoram P, Molitor J, Plagnol V, Tavar{\'e} S: \textbf{Markov Chain {M}onte
  {C}arlo Without Likelihoods}. \emph{Proc Natl Acad Sci USA} 2003,
  \textbf{100}(26):15324--15328.

\bibitem{beaumont08}
Beaumont MA: \textbf{Joint determination of topology, divergence time, and
  immigration in population trees}. In \emph{Simulation, Genetics and Human
  Prehistory}, McDonald Institute Monographs. Edited by Matsumura~S RC
  Forster~P, UK: McDonald Institute Monographs 2008:134--1541.

\bibitem{francoisetal08}
Fran\c{c}ois O, Blum MGB, Jakobsson M, Rosenberg NA: \textbf{{Demographic
  History of European Populations of \emph{{A}rabidopsis thaliana}}}.
  \emph{PLoS Genet} 2008, \textbf{4}:e1000075.

\bibitem{francoislaval11}
Fran{\c c}ois O, Laval G: \textbf{Deviance Information Criteria for Model
  Selection in Approximate Bayesian Computation}. \emph{arXiv} 2011,
  \urlprefix\url{[http://arxiv.org/abs/1105.0269]}.

\bibitem{hastieetal03}
Hastie T, Tibshirani R, Friedman JH: \emph{{The Elements of Statistical
  Learning}}. Springer, 2nd edition 2009.

\bibitem{voightetal05}
Voight BF, Adams AM, Frisse LA, Qian Y, Hudson RR, {Di Rienzo} A:
  \textbf{Interrogating multiple aspects of variation in a full resequencing
  data set to infer human population size changes}. \emph{Proc Natl Acad Sci
  USA} 2005, \textbf{102}:18508--18513.

\bibitem{ratmannetal09}
Ratmann O, Andrieu C, Wiuf C, Richardson S: \textbf{Model criticism based on
  likelihood-free inference, with an application to protein network evolution}.
  \emph{Proceedings of the National Academy of Sciences} 2009,
  \textbf{106}(26):10576--10581,
  \urlprefix\url{[http://www.pnas.org/content/106/26/10576.abstract]}.

\end{thebibliography}

\newcommand{\BMCxmlcomment}[1]{}

\BMCxmlcomment{

<refgrp>

<bibl id="B1">
  <title><p>Approximate {B}ayesian computation in population
  genetics</p></title>
  <aug>
    <au><snm>Beaumont</snm><fnm>M A</fnm></au>
    <au><snm>Zhang</snm><fnm>W</fnm></au>
    <au><snm>Balding</snm><fnm>D J</fnm></au>
  </aug>
  <source>Genetics</source>
  <pubdate>2002</pubdate>
  <volume>162</volume>
  <fpage>2025</fpage>
  <lpage>2035</lpage>
</bibl>

<bibl id="B2">
  <title><p>Population growth of human {Y} chromosomes: a study of Y chromosome
  microsatellites</p></title>
  <aug>
    <au><snm>Pritchard</snm><fnm>J K</fnm></au>
    <au><snm>Seielstad</snm><fnm>M T</fnm></au>
    <au><snm>Perez Lezaun</snm><fnm>A</fnm></au>
    <au><snm>Feldman</snm><fnm>M W</fnm></au>
  </aug>
  <source>Molecular Biology and Evolution</source>
  <pubdate>1999</pubdate>
  <volume>16</volume>
  <fpage>1791</fpage>
  <lpage>1798</lpage>
</bibl>

<bibl id="B3">
  <title><p>Approximate Bayesian Computation in Evolution and
  Ecology</p></title>
  <aug>
    <au><snm>Beaumont</snm><fnm>MA</fnm></au>
  </aug>
  <source>Annual Review of Ecology, Evolution, and Systematics</source>
  <pubdate>2010</pubdate>
  <volume>41</volume>
  <fpage>379</fpage>
  <lpage>406</lpage>
</bibl>

<bibl id="B4">
  <title><p>ABC as a flexible framework to estimate demography over space and
  time: some cons, many pros</p></title>
  <aug>
    <au><snm>Bertorelle</snm><fnm>G.</fnm></au>
    <au><snm>Benazzo</snm><fnm>A.</fnm></au>
    <au><snm>Mona</snm><fnm>S.</fnm></au>
  </aug>
  <source>Molecular Ecology</source>
  <pubdate>2010</pubdate>
  <volume>19</volume>
  <fpage>2609</fpage>
  <lpage>-2625</lpage>
</bibl>

<bibl id="B5">
  <title><p>{Approximate Bayesian Computation in practice}</p></title>
  <aug>
    <au><snm>Csill\'ery</snm><fnm>K</fnm></au>
    <au><snm>Blum</snm><fnm>M G B</fnm></au>
    <au><snm>Gaggiotti</snm><fnm>O E</fnm></au>
    <au><snm>Fran{\c c}ois</snm><fnm>O</fnm></au>
  </aug>
  <source>Trends Ecol Evol</source>
  <pubdate>2010</pubdate>
  <volume>25</volume>
  <fpage>410</fpage>
  <lpage>418</lpage>
</bibl>

<bibl id="B6">
  <title><p>Serial SimCoal: A population genetics model for data from multiple
  populations and points in time</p></title>
  <aug>
    <au><snm>Anderson</snm><fnm>CNK</fnm></au>
    <au><snm>Ramakrishnan</snm><fnm>U</fnm></au>
    <au><snm>Chan</snm><fnm>YL</fnm></au>
    <au><snm>Hadly</snm><fnm>EA</fnm></au>
  </aug>
  <source>Bioinformatics</source>
  <pubdate>2005</pubdate>
  <volume>21</volume>
  <issue>8</issue>
  <fpage>1733</fpage>
  <lpage>-1734</lpage>
  <url>http://bioinformatics.oxfordjournals.org/content/21/8/1733.abstract</url>
</bibl>

<bibl id="B7">
  <title><p>{Inferring population history with DIY ABC: a user-friendly
  approach to approximate Bayesian computation}</p></title>
  <aug>
    <au><snm>Cornuet</snm><fnm>JM</fnm></au>
    <au><snm>Santos</snm><fnm>F</fnm></au>
    <au><snm>Beaumont</snm><fnm>MA</fnm></au>
    <au><snm>Robert</snm><fnm>CP</fnm></au>
    <au><snm>Marin</snm><fnm>JM</fnm></au>
    <au><snm>Balding</snm><fnm>DJ</fnm></au>
    <au><snm>Guillemaud</snm><fnm>T</fnm></au>
    <au><snm>Estoup</snm><fnm>A</fnm></au>
  </aug>
  <source>Bioinformatics</source>
  <pubdate>2008</pubdate>
  <volume>24</volume>
  <issue>23</issue>
  <fpage>2713</fpage>
  <lpage>2719</lpage>
  <url>http://bioinformatics.oxfordjournals.org/content/24/23/2713.abstract</url>
</bibl>

<bibl id="B8">
  <title><p>COMPUTER PROGRAMS: onesamp: a program to estimate effective
  population size using approximate Bayesian computation</p></title>
  <aug>
    <au><snm>Tallmon</snm><fnm>D A</fnm></au>
    <au><snm>Koyuk</snm><fnm>A</fnm></au>
    <au><snm>Luikart</snm><fnm>G</fnm></au>
    <au><snm>Beaumont</snm><fnm>M A</fnm></au>
  </aug>
  <source>Molecular Ecology Resources</source>
  <publisher>Blackwell Publishing Ltd</publisher>
  <pubdate>2008</pubdate>
  <volume>8</volume>
  <issue>2</issue>
  <fpage>299</fpage>
  <lpage>-301</lpage>
  <url>http://dx.doi.org/10.1111/j.1471-8286.2007.01997.x</url>
</bibl>

<bibl id="B9">
  <title><p>{PopABC: a program to infer historical demographic
  parameters}</p></title>
  <aug>
    <au><snm>Lopes</snm><fnm>JS</fnm></au>
    <au><snm>Balding</snm><fnm>D</fnm></au>
    <au><snm>Beaumont</snm><fnm>MA</fnm></au>
  </aug>
  <source>Bioinformatics</source>
  <pubdate>2009</pubdate>
  <volume>25</volume>
  <issue>20</issue>
  <fpage>2747</fpage>
  <lpage>2749</lpage>
  <url>http://bioinformatics.oxfordjournals.org/content/25/20/2747.abstract</url>
</bibl>

<bibl id="B10">
  <title><p>Automating approximate Bayesian computation by local linear
  regression</p></title>
  <aug>
    <au><snm>Thornton</snm><fnm>K</fnm></au>
  </aug>
  <source>BMC Genetics</source>
  <pubdate>2009</pubdate>
  <volume>10</volume>
  <issue>1</issue>
  <fpage>35</fpage>
  <url>http://www.biomedcentral.com/1471-2156/10/35</url>
</bibl>

<bibl id="B11">
  <title><p>Inference on population history and model checking using DNA
  sequence and microsatellite data with the software DIYABC (v1.0)</p></title>
  <aug>
    <au><snm>Cornuet</snm><fnm>JM</fnm></au>
    <au><snm>Ravigne</snm><fnm>V</fnm></au>
    <au><snm>Estoup</snm><fnm>A</fnm></au>
  </aug>
  <source>BMC Bioinformatics</source>
  <pubdate>2010</pubdate>
  <volume>11</volume>
  <issue>1</issue>
  <fpage>401</fpage>
  <url>http://www.biomedcentral.com/1471-2105/11/401</url>
</bibl>

<bibl id="B12">
  <title><p>{ABC-SysBio---approximate Bayesian computation in Python with GPU
  support}</p></title>
  <aug>
    <au><snm>Liepe</snm><fnm>J</fnm></au>
    <au><snm>Barnes</snm><fnm>C</fnm></au>
    <au><snm>Cule</snm><fnm>E</fnm></au>
    <au><snm>Erguler</snm><fnm>K</fnm></au>
    <au><snm>Kirk</snm><fnm>P</fnm></au>
    <au><snm>Toni</snm><fnm>T</fnm></au>
    <au><snm>Stumpf</snm><fnm>MP</fnm></au>
  </aug>
  <source>Bioinformatics</source>
  <pubdate>2010</pubdate>
  <volume>26</volume>
  <issue>14</issue>
  <fpage>1797</fpage>
  <lpage>1799</lpage>
  <url>http://bioinformatics.oxfordjournals.org/content/26/14/1797.abstract</url>
</bibl>

<bibl id="B13">
  <title><p>ABCtoolbox: a versatile toolkit for approximate Bayesian
  computations</p></title>
  <aug>
    <au><snm>Wegmann</snm><fnm>D</fnm></au>
    <au><snm>Leuenberger</snm><fnm>C</fnm></au>
    <au><snm>Neuenschwander</snm><fnm>S</fnm></au>
    <au><snm>Excoffier</snm><fnm>L</fnm></au>
  </aug>
  <source>BMC Bioinformatics</source>
  <pubdate>2010</pubdate>
  <volume>11</volume>
  <issue>1</issue>
  <fpage>116</fpage>
  <url>http://www.biomedcentral.com/1471-2105/11/116</url>
</bibl>

<bibl id="B14">
  <title><p>2BAD: an application to estimate the parental contributions during
  two independent admixture events</p></title>
  <aug>
    <au><snm>Bray</snm><fnm>T C</fnm></au>
    <au><snm>Sousa</snm><fnm>VC</fnm></au>
    <au><snm>Parreira</snm><fnm>B</fnm></au>
    <au><snm>Bruford</snm><fnm>MW</fnm></au>
    <au><snm>Chikhi</snm><fnm>L</fnm></au>
  </aug>
  <source>Molecular Ecology Resources</source>
  <pubdate>2010</pubdate>
  <volume>10</volume>
  <fpage>538,541</fpage>
</bibl>

<bibl id="B15">
  <title><p>msBayes: Pipeline for testing comparative phylogeographic histories
  using hierarchical approximate Bayesian computation</p></title>
  <aug>
    <au><snm>Hickerson</snm><fnm>M</fnm></au>
    <au><snm>Stahl</snm><fnm>E</fnm></au>
    <au><snm>Takebayashi</snm><fnm>N</fnm></au>
  </aug>
  <source>BMC Bioinformatics</source>
  <pubdate>2007</pubdate>
  <volume>8</volume>
  <issue>1</issue>
  <fpage>268</fpage>
  <url>http://www.biomedcentral.com/1471-2105/8/268</url>
</bibl>

<bibl id="B16">
  <title><p>MTML-msBayes: Approximate Bayesian comparative phylogeographic
  inference from multiple taxa and multiple loci with rate
  heterogeneity</p></title>
  <aug>
    <au><snm>Huang</snm><fnm>W</fnm></au>
    <au><snm>Takebayashi</snm><fnm>N</fnm></au>
    <au><snm>Qi</snm><fnm>Y</fnm></au>
    <au><snm>Hickerson</snm><fnm>M</fnm></au>
  </aug>
  <source>BMC Bioinformatics</source>
  <pubdate>2011</pubdate>
  <volume>12</volume>
  <issue>1</issue>
  <fpage>1</fpage>
  <url>http://www.biomedcentral.com/1471-2105/12/1</url>
</bibl>

<bibl id="B17">
  <title><p>REJECTOR: software for population history inference from genetic
  data via a rejection algorithm</p></title>
  <aug>
    <au><snm>Jobin</snm><fnm>MJ</fnm></au>
    <au><snm>Mountain</snm><fnm>JL</fnm></au>
  </aug>
  <source>Bioinformatics</source>
  <pubdate>2008</pubdate>
  <volume>24</volume>
  <fpage>2936</fpage>
  <lpage>2937</lpage>
</bibl>

<bibl id="B18">
  <title><p>Inference for stereological extremes</p></title>
  <aug>
    <au><snm>Bortot</snm><fnm>P</fnm></au>
    <au><snm>Coles</snm><fnm>S G</fnm></au>
    <au><snm>Sisson</snm><fnm>S A</fnm></au>
  </aug>
  <source>Journal of the American Statistical Association</source>
  <pubdate>2007</pubdate>
  <volume>102</volume>
  <fpage>84</fpage>
  <lpage>-92</lpage>
</bibl>

<bibl id="B19">
  <title><p>Sequential {M}onte {C}arlo without likelihoods</p></title>
  <aug>
    <au><snm>Sisson</snm><fnm>S A</fnm></au>
    <au><snm>Fan</snm><fnm>Y</fnm></au>
    <au><snm>Tanaka</snm><fnm>M</fnm></au>
  </aug>
  <source>Proceedings of the National Academy of Sciences of the United States
  of America</source>
  <pubdate>2007</pubdate>
  <volume>104</volume>
  <fpage>1760</fpage>
  <lpage>1765</lpage>
  <note>Errata (2009), 106, 16889</note>
</bibl>

<bibl id="B20">
  <title><p>Adaptivity for {A}{B}{C} algorithms: the {A}{B}{C}-{P}{M}{C}
  scheme</p></title>
  <aug>
    <au><snm>Beaumont</snm><fnm>M A</fnm></au>
    <au><snm>Marin</snm><fnm>J M</fnm></au>
    <au><snm>Cornuet</snm><fnm>J M</fnm></au>
    <au><snm>Robert</snm><fnm>C P</fnm></au>
  </aug>
  <source>Biometrika</source>
  <pubdate>2009</pubdate>
  <volume>96</volume>
  <fpage>983</fpage>
  <lpage>990</lpage>
</bibl>

<bibl id="B21">
  <title><p>Approximate Bayesian Computation: A Nonparametric
  Perspective</p></title>
  <aug>
    <au><snm>Blum</snm><fnm>MGB</fnm></au>
  </aug>
  <source>Journal of the American Statistical Association</source>
  <pubdate>2010</pubdate>
  <volume>105</volume>
  <issue>491</issue>
  <fpage>1178</fpage>
  <lpage>1187</lpage>
  <url>http://pubs.amstat.org/doi/abs/10.1198/jasa.2010.tm09448</url>
</bibl>

<bibl id="B22">
  <title><p>R: A Language and Environment for Statistical Computing</p></title>
  <aug>
    <au><cnm>{R Development Core Team}</cnm></au>
  </aug>
  <publisher>Vienna, Austria</publisher>
  <pubdate>2010</pubdate>
  <url>http://www.R-project.org</url>
  <note>{ISBN} 3-900051-07-0</note>
</bibl>

<bibl id="B23">
  <title><p>{Generating samples under a Wright-Fisher neutral model of genetic
  variation}</p></title>
  <aug>
    <au><snm>Hudson</snm><fnm>RR</fnm></au>
  </aug>
  <source>Bioinformatics</source>
  <pubdate>2002</pubdate>
  <volume>18</volume>
  <issue>2</issue>
  <fpage>337</fpage>
  <lpage>338</lpage>
</bibl>

<bibl id="B24">
  <title><p>{SIMCOAL} 2.0: a program to simulate genomic diversity over large
  recombining regions in a subdivided population with a complex
  history</p></title>
  <aug>
    <au><snm>Laval</snm><fnm>G</fnm></au>
    <au><snm>Excoffier</snm><fnm>L</fnm></au>
  </aug>
  <source>Bioinformatics</source>
  <pubdate>2004</pubdate>
  <volume>20</volume>
  <issue>15</issue>
  <fpage>2485</fpage>
  <lpage>-2487</lpage>
</bibl>

<bibl id="B25">
  <title><p>Bayesian Data Analysis, Second Edition (Texts in Statistical
  Science)</p></title>
  <aug>
    <au><snm>Gelman</snm><fnm>A</fnm></au>
    <au><snm>Carlin</snm><fnm>JB</fnm></au>
    <au><snm>Stern</snm><fnm>HS</fnm></au>
    <au><snm>Rubin</snm><fnm>DB</fnm></au>
  </aug>
  <source>Hardcover</source>
  <publisher>Boca Raton: Chapman \& Hall/CRC</publisher>
  <edition>2</edition>
  <pubdate>2003</pubdate>
</bibl>

<bibl id="B26">
  <title><p>Non-linear regression models for {A}pproximate {B}ayesian
  {C}omputation</p></title>
  <aug>
    <au><snm>Blum</snm><fnm>M G B</fnm></au>
    <au><snm>Fran\c{c}ois</snm><fnm>O</fnm></au>
  </aug>
  <source>Statistics and Computing</source>
  <pubdate>2010</pubdate>
  <volume>20</volume>
  <fpage>63</fpage>
  <lpage>73</lpage>
</bibl>

<bibl id="B27">
  <title><p>Adaptive approximate Bayesian computation</p></title>
  <aug>
    <au><snm>Beaumont</snm><fnm>MA</fnm></au>
    <au><snm>Cornuet</snm><fnm>JM</fnm></au>
    <au><snm>Marin</snm><fnm>JM</fnm></au>
    <au><snm>Robert</snm><fnm>CP</fnm></au>
  </aug>
  <source>Biometrika</source>
  <pubdate>2009</pubdate>
  <volume>96</volume>
  <issue>4</issue>
  <fpage>983</fpage>
  <lpage>-990</lpage>
</bibl>

<bibl id="B28">
  <title><p>Markov Chain {M}onte {C}arlo Without Likelihoods</p></title>
  <aug>
    <au><snm>Marjoram</snm><fnm>P</fnm></au>
    <au><snm>Molitor</snm><fnm>J</fnm></au>
    <au><snm>Plagnol</snm><fnm>V</fnm></au>
    <au><snm>Tavar{\'e}</snm><fnm>S</fnm></au>
  </aug>
  <source>Proc Natl Acad Sci USA</source>
  <pubdate>2003</pubdate>
  <volume>100</volume>
  <issue>26</issue>
  <fpage>15324</fpage>
  <lpage>-15328</lpage>
</bibl>

<bibl id="B29">
  <title><p>Joint determination of topology, divergence time, and immigration
  in population trees</p></title>
  <aug>
    <au><snm>Beaumont</snm><fnm>M A</fnm></au>
  </aug>
  <source>Simulation, Genetics and Human Prehistory</source>
  <publisher>UK: McDonald Institute Monographs</publisher>
  <editor>Matsumura S, Forster P, Renfrew C</editor>
  <series><title><p>McDonald Institute Monographs</p></title></series>
  <pubdate>2008</pubdate>
  <fpage>134</fpage>
  <lpage>-1541</lpage>
</bibl>

<bibl id="B30">
  <title><p>{Demographic History of European Populations of \emph{{A}rabidopsis
  thaliana}}</p></title>
  <aug>
    <au><snm>Fran\c{c}ois</snm><fnm>O</fnm></au>
    <au><snm>Blum</snm><fnm>MGB</fnm></au>
    <au><snm>Jakobsson</snm><fnm>M</fnm></au>
    <au><snm>Rosenberg</snm><fnm>NA</fnm></au>
  </aug>
  <source>PLoS Genet</source>
  <pubdate>2008</pubdate>
  <volume>4</volume>
  <fpage>e1000075</fpage>
</bibl>

<bibl id="B31">
  <title><p>Deviance Information Criteria for Model Selection in Approximate
  Bayesian Computation</p></title>
  <aug>
    <au><snm>Fran{\c c}ois</snm><fnm>O</fnm></au>
    <au><snm>Laval</snm><fnm>G</fnm></au>
  </aug>
  <source>arXiv</source>
  <pubdate>2011</pubdate>
  <url>http://arxiv.org/abs/1105.0269</url>
</bibl>

<bibl id="B32">
  <title><p>{The Elements of Statistical Learning}</p></title>
  <aug>
    <au><snm>Hastie</snm><fnm>T.</fnm></au>
    <au><snm>Tibshirani</snm><fnm>R.</fnm></au>
    <au><snm>Friedman</snm><fnm>J. H.</fnm></au>
  </aug>
  <publisher>Springer</publisher>
  <edition>2</edition>
  <pubdate>2009</pubdate>
</bibl>

<bibl id="B33">
  <title><p>Interrogating multiple aspects of variation in a full resequencing
  data set to infer human population size changes</p></title>
  <aug>
    <au><snm>Voight</snm><fnm>B. F.</fnm></au>
    <au><snm>Adams</snm><fnm>A. M.</fnm></au>
    <au><snm>Frisse</snm><fnm>L. A.</fnm></au>
    <au><snm>Qian</snm><fnm>Y.</fnm></au>
    <au><snm>Hudson</snm><fnm>R. R.</fnm></au>
    <au><snm>{Di Rienzo}</snm><fnm>A.</fnm></au>
  </aug>
  <source>Proc Natl Acad Sci USA</source>
  <pubdate>2005</pubdate>
  <volume>102</volume>
  <fpage>18508</fpage>
  <lpage>18513</lpage>
</bibl>

<bibl id="B34">
  <title><p>Model criticism based on likelihood-free inference, with an
  application to protein network evolution</p></title>
  <aug>
    <au><snm>Ratmann</snm><fnm>O</fnm></au>
    <au><snm>Andrieu</snm><fnm>C</fnm></au>
    <au><snm>Wiuf</snm><fnm>C</fnm></au>
    <au><snm>Richardson</snm><fnm>S</fnm></au>
  </aug>
  <source>Proceedings of the National Academy of Sciences</source>
  <pubdate>2009</pubdate>
  <volume>106</volume>
  <issue>26</issue>
  <fpage>10576</fpage>
  <lpage>-10581</lpage>
  <url>http://www.pnas.org/content/106/26/10576.abstract</url>
</bibl>

</refgrp>
} 

\ifthenelse{\boolean{publ}}{\end{multicols}}{}


\newpage

\section*{Figures}

  \subsection*{Figure 1 - Model misclassification}
  Graphical illustration of the confusion matrix for three demographic
  models. Each color from dark to light gray corresponds to a model
  (bottleneck, constant, exponential, respectively). If the simulations were perfectly classified, each bar would have a single color of its own corresponding model.

  \subsection*{Figure 2 - Posterior predictive checks}
   Distribution of the three summary statistics a posteriori for the
   three demographic models. Vertical bars correspond to the values
   computed from the Italian data.
 
  \subsection*{Figure 3 - Cross-validation for parameter estimation}
   Estimated values as a function of true parameter values of the
   ancestral population size, $N_a$, under the bottleneck model.  The
   plot has been generated by the function \verb|plot| from an object
   of class \verb|"cv4abc"|. Point estimates are obtained using
   posterior median.
 
  \subsection*{Figure 4 - ABC regression diagnostics}
   Diagnostic plot of an object of class \verb|"abc"|, generated by
   the function \verb|plot|. The upper left panel shows the prior
   distribution. The lower left panel shows the posterior distribution
   obtained with and without the regression correction method, and the
   prior distribution, for reference. The upper right panel displays
   the distances between observed and simulated summary statistics as
   a function of the parameter values. Red points indicate the
   accepted values. The lower right panel is a Normal Q-Q plot of the
   residuals of the regression, here from \verb|nnet|.

\clearpage


\section*{Tables}
\subsection*{Table 1 - Posterior probabilities of the different demographic models}
\par \mbox{}
\par
\mbox{
  \begin{tabular}{c|cc|ccc}
    Population & Method  & Tolerance rate & Constant & Expansion & Bottleneck\\
    \hline
    &   & 0.001 &  0.28 & 0.61 & 0.11 \\
    & Regression &                  0.005 &  0.35 & 0.52 & 0.13\\
    Hausa	 &  &                  0.01 &   0.34 & 0.54 & 0.12 \\ 
    \cline{2-6}
    (Africa)	 &         & 0.001 &  0.27 & 0.58 & 0.15 \\
    & Rejection &                  0.005 &  0.35 & 0.52 & 0.13\\
    &  &                  0.01  &  0.33 & 0.53 & 0.14\\
    \hline
    &  & 0.001 &  0.24 & 0.01 & 0.75\\
    & Regression  &                   0.005 &  0.20 & 0.01 & 0.79\\
    Italian & &                   0.01  &  0.21 & 0.01 & 0.78\\
    \cline{2-6}
    (Europe)  &                   & 0.001 &  0.31 & 0.03 & 0.66\\
    & Rejection &                   0.005 &  0.36 & 0.05 & 0.59\\
    & &                   0.01  &  0.39 & 0.05 & 0.56\\
    \hline
  \end{tabular}
}

\clearpage
\newpage

\begin{center}
\includegraphics[width=10cm,angle=0]{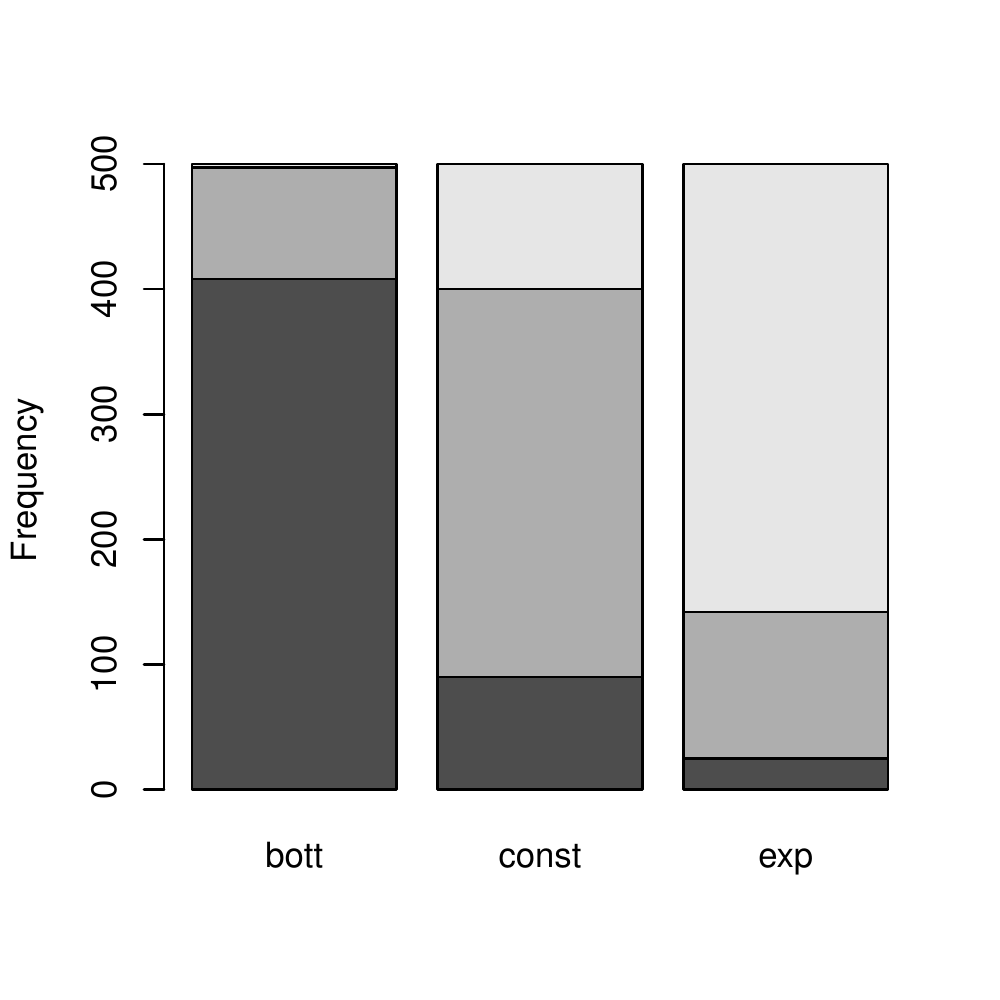}
\end{center}

\clearpage
\newpage

\begin{center}
\includegraphics[width=10cm,angle=0]{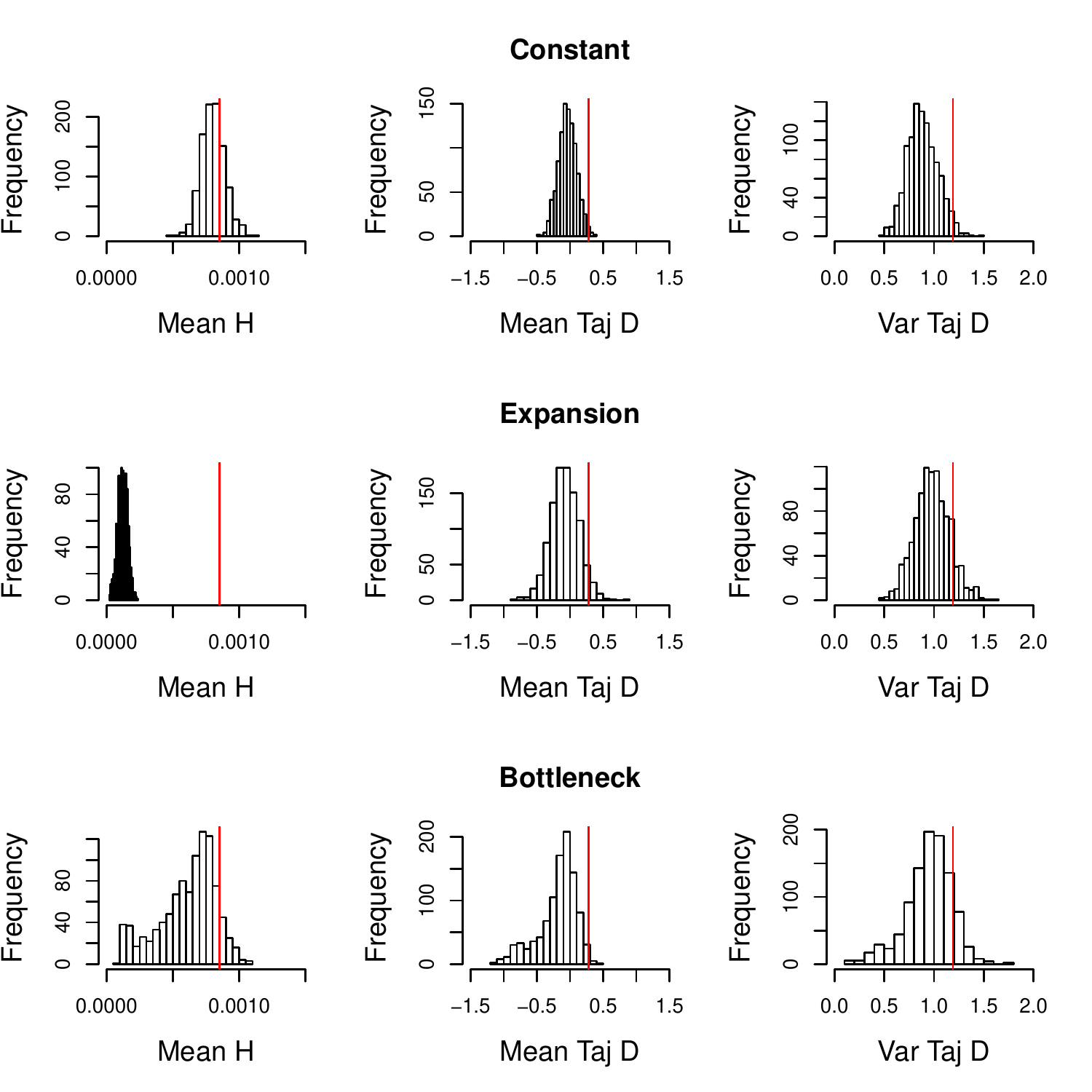}
\end{center}

\clearpage
\newpage

\begin{center}
\includegraphics[width=10cm,angle=0]{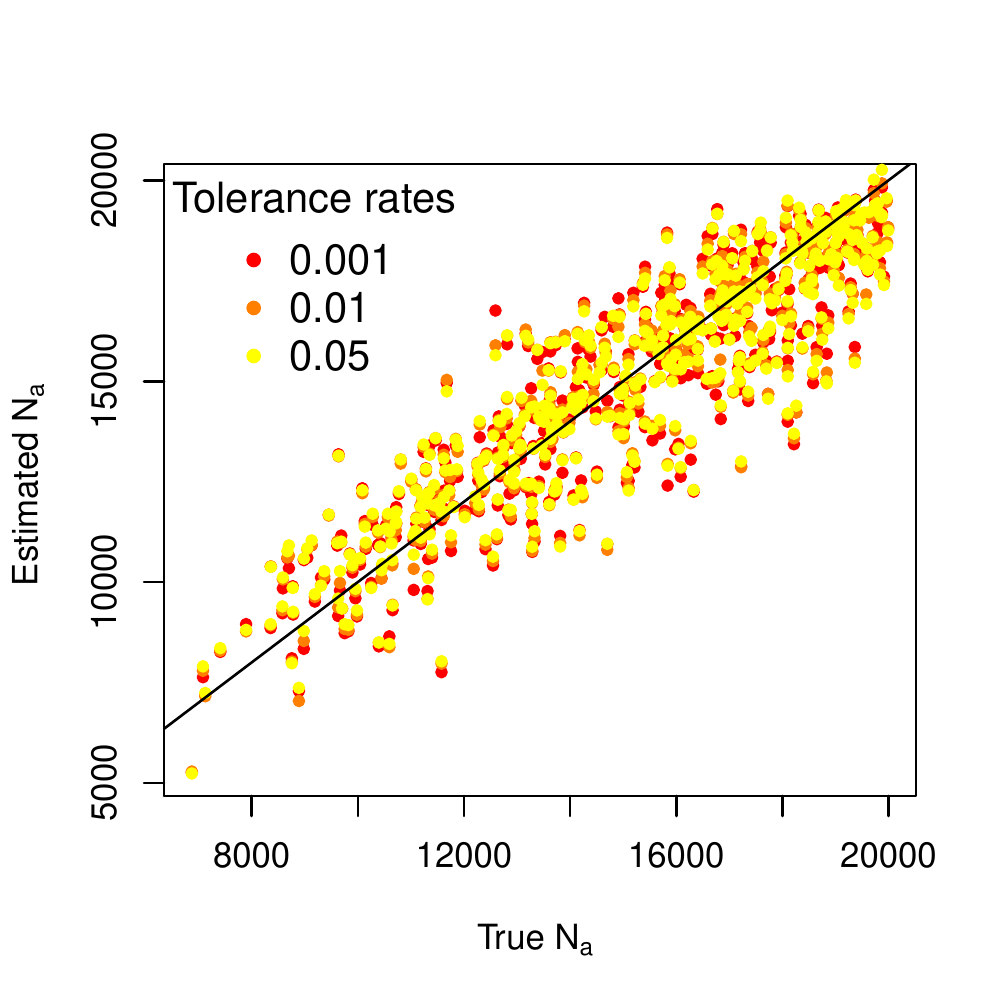}
\end{center}

\clearpage
\newpage

\begin{center}
\includegraphics[width=10cm,angle=0]{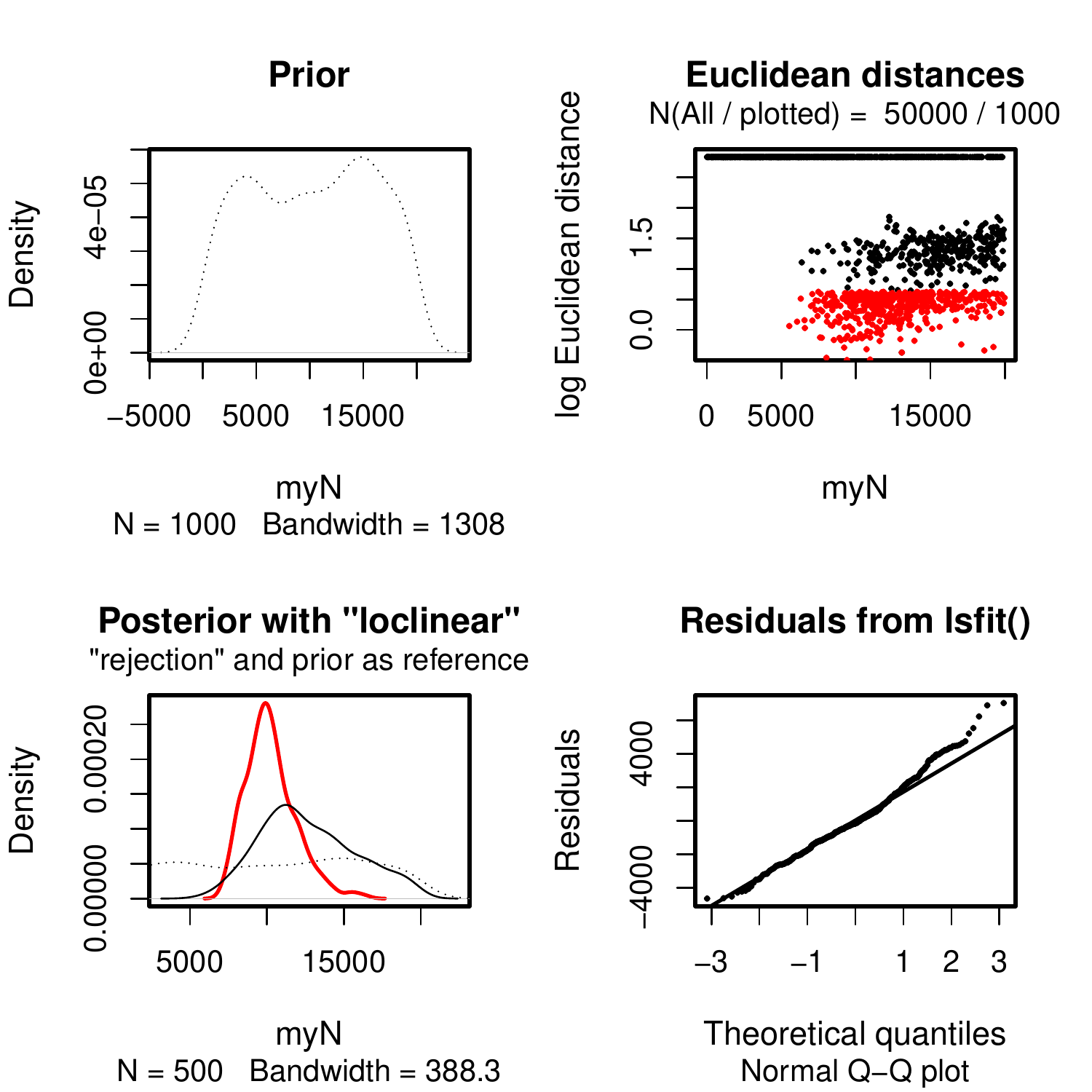}
\end{center}

\end{bmcformat}
\end{document}